\RequirePackage{fix-cm}
\documentclass[twocolumn]{svjour3}          
\smartqed  
\usepackage{graphicx,amssymb}
\begin{document}

\title{Physics Criteria for a Subscale Plasma Liner Experiment\thanks{This work was
supported in part by the U.S. Department of Energy under contract no.\ DE-AC52-06NA25396.
}}


\author{Scott C. Hsu \and Y. C. Francis Thio}


\institute{Scott C. Hsu \at
Los Alamos National Laboratory, Los Alamos, NM 87545 \\
Tel.: +1-505-667-3386\\
\email{scotthsu@lanl.gov} 
\and
Y. C. Francis Thio \at
HyperJet Fusion Corporation, Chantilly, VA  20151\\
Tel:  +1-301-524-4698\\
\email{francis.thio@hyperjetfusion.com}
}         
              
\date{Received: date / Accepted: date}

\maketitle

\begin{abstract}
Spherically imploding plasma liners,
formed by merging hypersonic plasma jets, are a proposed standoff
driver to compress magnetized target plasmas to fusion conditions [S. C. Hsu et al., IEEE Trans.\ Plasma Sci.~{\bf 40}, 1287 (2012)].  In
this paper, the parameter space and 
physics criteria are identified for a subscale, plasma-liner-formation experiment to provide data,
e.g., on liner ram-pressure scaling and uniformity, 
that are relevant for addressing scientific issues
of full-scale plasma liners required to achieve fusion conditions.
Based on these criteria,
we quantitatively estimate the minimum liner kinetic energy and mass needed, which
informed the design of a subscale plasma liner experiment now under development.
\end{abstract}

\section{Introduction}
\label{sec:intro}

Ongoing research \cite{hsu18} on the Plasma Liner Experiment (PLX) \cite{hsu12pop,hsu15jpp}
is aiming to demonstrate
the formation and implosion of spherical plasma liners via merging hypersonic plasma jets
formed by pulsed coaxial guns (where ``hypersonic'' refers to the jets being 
highly supersonic and where atomic excitation, ionization, and radiative
effects are important \cite{bertin13}).
The guns, jets, and imploding plasma liner constitute a {\em driver} to compress a magnetized plasma
{\em target} to fusion conditions, i.e., a proposed embodiment of magneto-inertial fusion (MIF) \cite{lindemuth83,kirkpatrick95,lindemuth09}
known as plasma-jet-driven magneto-inertial fusion
(PJMIF) \cite{thio99,thio01,hsu12ieee,knapp14}.  PJMIF has several attributes making it a potential candidate for an economically viable, repetitively pulsed fusion reactor \cite{coststudy17}.
An immediate, near-term objective is to retire the major physics risks associated with the plasma-liner-driver aspects of PJMIF at the lowest-possible cost and technical risk.

The key goal of PJMIF development during the present three-year research phase (ending in 2019) 
is to demonstrate the
formation, viability, and scalability of spherically imploding plasma liners formed
by merging plasma jets.  Specifically, an objective
is to obtain experimental data on two key scientific issues of the plasma liner as
an MIF compression driver:  (1)~scaling of peak
ram pressure ($\rho v^2$) of the plasma liner versus initial plasma jet parameters
and number of jets, and (2)~evolution and control of non-uniformities seeded by the jet-merging process,
which have the potential to degrade the ability of a plasma liner to compress a target
plasma to fusion conditions.

Because the overall cost of a plasma-liner-formation experiment
is closely linked to the initial stored energy, we were motivated to
conduct a careful analysis of the minimum stored energy required to address
PJMIF-relevant, plasma-liner issues in a meaningful way.  
The required stored energy is determined by two independent properties:  (1)~the
electrical efficiency of the plasma guns, and (2)~the required minimum initial kinetic energy of the
imploding plasma liner.  This paper is restricted to consideration of the latter.
The analysis is based on
first identifying key physics criteria that must be satisfied in order for a subscale experiment to
provide data that is relevant for a full-scale, fusion-relevant plasma liner.   Based on these criteria,
we then quantitatively estimate the minimum initial plasma-liner kinetic
energy and mass of a subscale experiment that satisfies the fusion-relevant plasma-liner criteria.

A full-scale PJMIF plasma liner  \cite{thio99,hsu12ieee}
would consist of an array of, perhaps, hundreds of
coaxial plasma guns uniformly mounted around a spherical chamber that is several meters in radius.  
It is envisioned  \cite{hsu12ieee} that a small subset of the guns fires first to form a
magnetized plasma target, followed immediately
by the remainder of the guns firing to form a spherically imploding plasma liner that compresses
the target.
The physical processes and steps of plasma-liner formation via merging plasma jets,
the subsequent convergence of the liner, and scalings of peak liner ram pressure and 
non-uniformity evolution have been studied extensively and presented elsewhere \cite{thio99,thio01,parks08,samulyak10,awe11,hsu12ieee,davis12,hsu12pop,cassibry12,kim12,merritt13,cassibry13,kim13,merritt14,langendorf17pop}, though much more research is needed to
validate the PJMIF concept.

A key issue for PJMIF is the required/achievable symmetry of the imploding plasma liner; 
this is being addressed in ongoing research \cite{hsu18}.  Prior numerical studies
\cite{cassibry12,kim13} employed 3D simulations to elucidate
the origin and evolution of non-uniformities seeded by shocks that
form between discrete merging jets.
Results based on 3D smooth-particle hydrodynamic (SPH) simulations by Cassibry et 
al.~\cite{cassibry12} indicated that 
late-time uniformity of a plasma liner formed with discrete jets was similar to that of an initially 
spherically symmetric liner, which is a favorable result.
However, further simulations are needed to explore the effects of spatial resolution
(i.e., number of particles in the simulation) and the value of artificial viscosity on the 
liner symmetry and its evolution.
Kim et al.~\cite{kim13}, using a 3D hydrodynamics code, 
provided a physical picture of
plasma-liner formation via merging plasma jets and liner-uniformity evolution.
These simulations predicted that ``primary'' shocks would form between adjacent
merging jets, and then the shocked plasmas associated with the ``primary shocks'' would merge to
form ``secondary shocks.''
This physical picture has been verified in experiments \cite{hsu18}.
However, in these parameter regimes, shock strength is over-predicted in single-fluid hydrodynamics
codes.  Further benchmarking studies,
along with new 3D simulations of plasma-liner formation
via the merging of up to hundreds of plasma jets, will be reported elsewhere.
Indeed, a key objective of ongoing research \cite{hsu18} is to work toward
setting requirements on and identifying limits of achievable liner uniformity.

Another key issue for PJMIF is the need to develop compatible targets that take advantage
of the high implosion speed ($>50$~km/s) of a spherically imploding plasma liner.
Some discussions of PJMIF-relevant target formation have appeared elsewhere
\cite{ryutov09,hsu12ieee,welch14}, but much further research on PJMIF-compatible
target development is needed.  As described by D.~Ryutov~\cite{ryutov09}, an interesting target for
compression is a high-$\beta$ (i.e., $\beta \gg 1$) object with a ``tangled field'' that,
simultaneously, allows for reduced cross-field thermal transport (i.e., Hall parameter
$\omega\tau \sim 3$) and adequate parallel electron thermal confinement time (due to the
long connection length of the tangled field).  Because $\beta \sim 10$, it is expected
that the issue of magnetohydrodynamic (MHD) instability would be sidestepped entirely because
there is not enough free energy in the magnetic field to instigate virulent global instabilities.
D.~Ryutov \cite{ryutov09} lays out detailed requirements for such a target to satisfy the above
properties but recognizes that creation of such a target plasma ``may not be a simple task.''
He also states that ``an intuitively appealing
way for creating such a target would be the use of numerous plasma guns
generating small-scale, magnetized plasma bunches and injection of such bunches into
a limited volume.''  This is precisely the plan which we intend to pursue in the near future.
The high implosion speed ($>50$~km/s) of a spherically imploding plasma liner,
enabling the compression of a several-cm-radius plasma target in $\sim 1$~$\mu$s, is
what enables the possibility of near-adiabatic heating of such a novel, high-$\beta$ 
plasma target.

The remainder of the paper is organized as follows.  Section~\ref{sec:calcs} presents PJMIF 
reactor-relevant plasma-liner parameters in order to identify the relevant physics criteria
for a subscale experiment.  Section~\ref{sec:criteria} concisely states these criteria.
Section~\ref{sec:requirements} uses the criteria to derive the minimum liner kinetic energy
and mass for a relevant subscale experiment.
Finally, Sec.~\ref{sec:summary} summarizes the main results of the paper.

\section{PJMIF reactor-relevant plasma-liner parameters}
\label{sec:calcs}

We start by considering the PJMIF reactor-relevant parameter regime.
Table~\ref{table:params} summarizes the desired physical parameters of the target at stagnation and
the imploding spherical plasma liner at the time of peak $\rho v^2$.
The peak target thermal pressure $p$ at stagnation is limited by the peak liner $\rho v^2$.
One-dimensional radiation-hydrodynamic simulations aimed
at reaching the conditions in Table~\ref{table:params} have shown fusion energy gains up to 30 \cite{knapp14}, where the
gain is defined as the fusion energy divided by the initial liner kinetic energy.
Similar cases were studied further using a 1D semi-analytic model of PJMIF \cite{langendorf17pop}.
These promising configurations were originally identified through hundreds of 1D simulations
by one of the authors.

\begin{table*}
\caption{\label{table:params}Summary of reference PJMIF target parameters at peak compression and 
liner parameters at the time of peak $\rho v^2$ just prior to peak compression.  
The $M$ estimate assumes $T_e=T_i=2$~eV, $Z=1$, and $\gamma=1.3$.}
\begin{center}
\begin{tabular}{lcccc}
\hline\noalign{\smallskip}
parameter &  target (DT) & liner (Xe) & liner (Kr) & liner (Ar)\\
\noalign{\smallskip}\hline\noalign{\smallskip}
pressure (Mbar) & $p=150$ & $\rho v^2=150$ & $\rho v^2=150$ & $\rho v^2=150$\\
mass density $\rho$ (g/cm$^3$) & 0.02 & 3.06 & 3.06 & 3.06\\
ion density $n_i$~(cm$^{-3}$) & $5\times 10^{21}$ & $1.40\times 10^{22}$ & $2.19\times 10^{22}$ & $4.59 \times10^{22}$ \\
velocity $v$ (km/s) & 0 & 70 & 70 & 70\\
compressed target radius $r_{\rm stag}$ (cm) & 0.4 & & &\\
Mach number  $M$ 	& & 36 & 29 & 20\\
\noalign{\smallskip}\hline
\end{tabular}
\end{center}
\end{table*}

The required plasma-liner conditions at the time of peak $\rho v^2$
dictate the plasma-jet initial conditions as well as the range of plasma-liner parameters 
over the entire implosion from the jet-merging radius $r_m$ down
to the stagnation radius $r_{\rm stag}$,
which are connected through the following relations,
\begin{equation}
\rho_0 \sim \rho_{\rm stag}\left(\frac{r_{\rm stag}}{r_m}\right)^2
\label{eq:n0}
\end{equation}
and
\begin{equation}
M_{\rm liner}=\frac{2E_0}{v^2}\approx 4\pi r_m^2 L \rho_0,
\label{eq:mass}
\end{equation}
where $\rho_0=n_0 m$ and $n_0$ are the liner mass density
and ion number density at $r_m$, respectively,
$\rho_{\rm stag}$ the liner mass density just prior to reaching $r_{\rm stag}$,
$M_{\rm liner}$ the total liner mass, 
$E_0$ the liner kinetic energy at $r_m$, $v$ the liner implosion speed (assumed to be constant
from $r_m$ to just before $r_{\rm stag}$), $L$ the liner thickness at $r_m$, and
$m$ the atomic mass of the liner species.
Note that Eq.~(\ref{eq:n0}) holds strictly only for a 1D steady-state convergent flow with constant $v$ and 
constant polytropic index $\gamma$.  In a real plasma-liner system, these assumptions are not
expected to hold exactly; nevertheless, Eq.~(\ref{eq:n0}) provides an adequate method for 
our present need to obtain an approximate relationship between $\rho_0$ and $\rho_{\rm stag}$.

\begin{figure}
\centerline{\includegraphics[width=2.5truein]{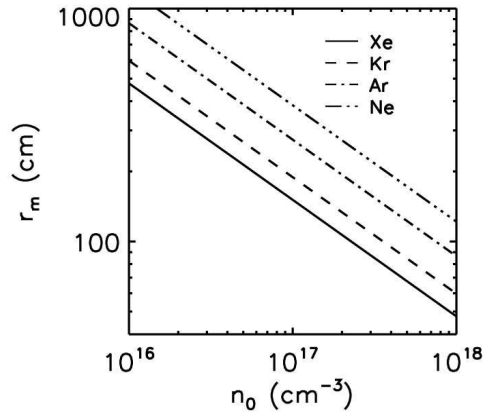}}
\caption{\label{fig:n0-rm}Liner merging radius $r_m$ vs.\ liner ion density $n_0$ at $r_m$, satisfying 
$\rho_0 r_m^2=0.50$~g/cm, for Xe, Kr, Ar, and Ne.}
\end{figure}

Using $\rho_{\rm stag}=3.06$~g/cm$^3$
and $r_{\rm stag}=0.4$~cm
from Table~\ref{table:params},
then Eq.~(\ref{eq:n0}) gives $\rho_0 r_m^2=0.50$~g/cm.
Using the latter and setting $E_0=25$~MJ, then $M_{\rm liner}=1.02\times 10^{-2}$~kg and $L=1.62$~cm.
Equation~(\ref{eq:mass}) shows that $L$ is fixed
once $\rho_{\rm stag} r_{\rm stag}^2 (= \rho_0 r_m^2)$, $E_0$, and $v$ are fixed.
Figure~\ref{fig:n0-rm} plots $r_m$ versus $n_0$ for Xe, Kr, Ar, and Ne (ion-to-proton 
mass ratios $\mu=131.29$, 83.80, 39.95, and
20.17, respectively), showing that we are limited to Xe and Kr liners if we restrict ourselves to
$r_m \lesssim 2$~m and $n_0 \lesssim 10^{17}$~cm$^{-3}$.

We proceed with estimating the important plasma, equation-of-state (EOS), and hydrodynamic
quantities of this reference PJMIF scenario.  We focus on the use of xenon in assessing the PJMIF
reactor-relevant regime because xenon gives the highest
$\rho_0$ for a given $n_0$ and the highest Mach number $M\equiv v/C_s$
(where $C_s\equiv [\gamma k(ZT_e+T_i)/m]^{1/2}$ is the ion sound speed) for given $T$ and $v$.

\subsection{Liner plasma-physics properties}

\subsubsection{Collisionality}
\label{sec:collisionality}
\begin{figure}
\centerline{\includegraphics[width=2.6truein]{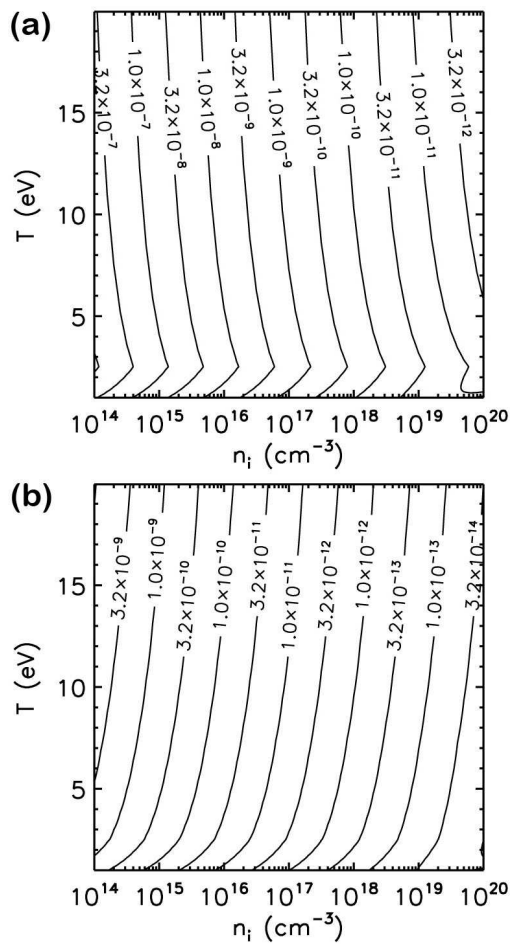}}
\caption{\label{fig:tau_c-Xe} Contours of thermal (a)~ion $\tau_i$ and (b) electron $\tau_e$ collision times (s) vs.\ $T$ ($=T_i=T_e$) and $n_i$
for xenon.  A simple model for ionization, $Z=0.63T^{1/2}$, is used \cite{drake10}.}
\end{figure}

The thermal ion and electron collision times, $\tau_i$ and $\tau_e$, respectively, in the plasma liner,
for a range of relevant $T$ ($=T_i=T_e$ assumed) and ion density
$n_i$ from $r_m$ down to $r_{\rm stag}$, are shown in Fig.~\ref{fig:tau_c-Xe} for  xenon.
The jogs in the $\tau_i\sim Z^{-4} n_i^{-1} T^{3/2}$ contours
are due to a transition from $Z=1$ ($\tau_i \sim n_i^{-1}T^{3/2}$) to
$Z\approx 0.63T^{1/2}$ ($\tau_i \sim n_i^{-1}T^{-1/2}$) when $T$ exceeds 2.5~eV (up to
a temperature at which xenon becomes fully stripped). 
The $Z\sim T^{1/2}$ relation is a convenient way to model increasing $Z$ with rising $T$ \cite{drake10}.
In contrast, $\tau_e \sim Z^{-1} n_i^{-1}T^{3/2}\sim n_i^{-1}T$ 
exhibits a milder jog at the same transition.
Figure~\ref{fig:tau_c-Xe} shows that the plasma liner is very collisional from $r_m$ 
down to $r_{\rm stag}$, with $\tau_i < 50$~ns and $\tau_e < 0.1$~ns for the entire evolution (assuming
that $n_i$ remains above $10^{16}$~cm$^{-3}$ and $T$ remains below 10~eV)\@.
The ion--electron energy equilibration time is $\tau_{Eie} \sim (m_i/m_e) \tau_e$.  For most of the liner
evolution, $\tau_{Eie} \ll \tau_{\rm implosion}\approx r_m/v \approx 21.4$~$\mu$s
(for $r_m=150$~cm and $v=70$~km/s), meaning that $T_e \approx
T_i$ is a good approximation.  The only exception is right at jet merging when the ions are shock 
heated over
a microsecond time scale when the liner is at relatively low density, in which case $\tau_{Eie} \sim 2.4$~$
\mu$s 
for $n\sim 10^{17}$~cm$^{-3}$ and $T \approx 5$--10~eV\@.  However, $T_e \approx T_i$ should likely be restored long before stagnation.  The degree of separation between $T_i$ and $T_e$
due to ion shock heating is the focus of ongoing
research \cite{hsu18} and will be reported further elsewhere.  The key conclusion here is that
a reactor-relevant plasma liner is highly collisional for its entire evolution.

\subsection{Liner equation-of-state and radiative properties}
\label{sec:eos-calcs}

\begin{figure}
\centerline{\includegraphics[width=3.2truein]{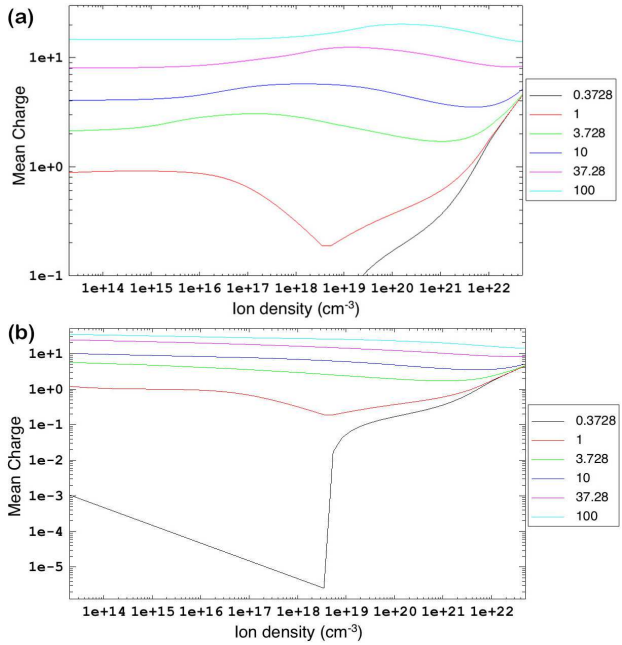}}
\caption{\label{fig:zbar-Xe}Mean-charge state $Z$
of xenon vs.\ ion density for different temperatures (in eV),
from (a)~non-LTE and (b)~LTE PROPACEOS calculations.  Differences between
the two models are pronounced at lower densities, indicating that
a non-LTE model should be used for the early stages of liner convergence.}%
\end{figure}

\begin{figure}
\centerline{\includegraphics[width=3.2truein]{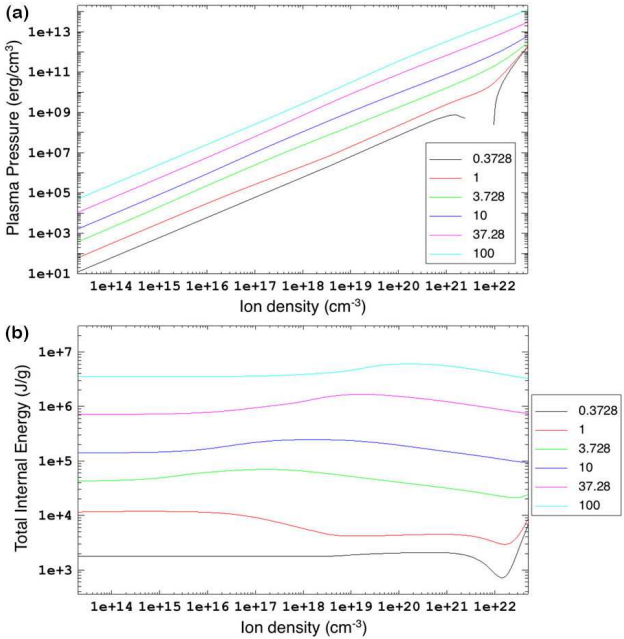}}
\caption{\label{fig:pressure-energy-Xe}(a)~Total thermal plasma
pressure and (b) internal energy (including
ion and electron thermal and ionization/excitation energies)
of xenon vs.\ ion density for different temperatures (in eV),
from a non-LTE PROPACEOS calculation.}%
\end{figure}

\begin{figure}
\centerline{\includegraphics[width=3.2truein]{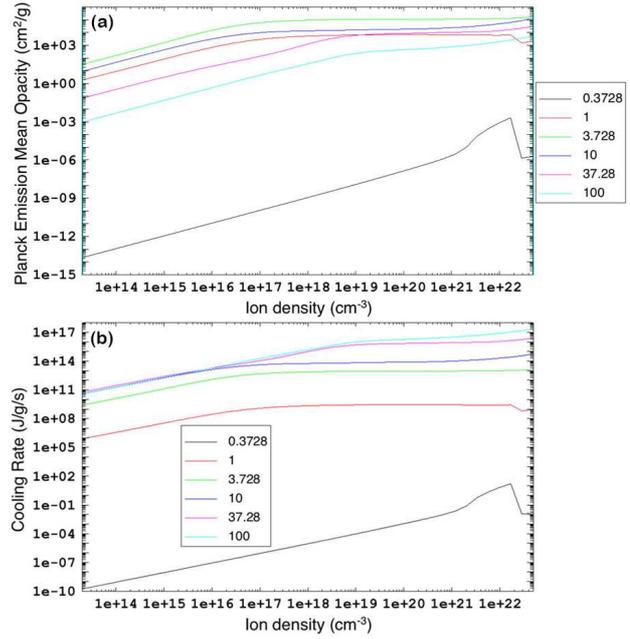}}
\caption{\label{fig:opacity-cooling-Xe}(a)~Frequency-integrated Planck emission mean opacity
and (b)~radiative-cooling rate of xenon vs.\ ion density for different temperatures (in eV),
from a non-LTE PROPACEOS calculation.}%
\end{figure}

\begin{figure}
\centerline{\includegraphics[width=2.8truein]{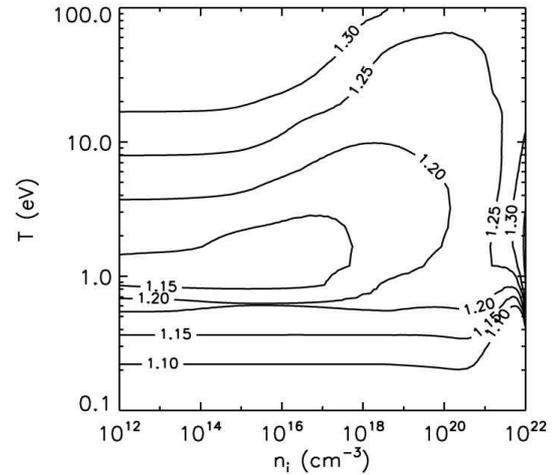}}
\caption{\label{fig:gamma-xe} Contours of polytropic index $\gamma$
as a function of $T$ ($=T_e=T_i$) and $n_i$,
calculated using Eq.~(\ref{eq:gamma}) and a non-LTE xenon EOS from PROPACEOS\@.}
\end{figure}

Non-local-thermodynamic-equilibrium (non-LTE) EOS 
and opacity tables are used to infer
the EOS (Figs.~\ref{fig:zbar-Xe}--\ref{fig:pressure-energy-Xe}),
opacity [Fig.~\ref{fig:opacity-cooling-Xe}(a)], and radiative-cooling
[Fig.~\ref{fig:opacity-cooling-Xe}(b)] properties of a reactor-relevant xenon plasma liner
for parameters spanning $r_m$ to $r_{\rm stag}$.  
The tables were generated using PROPACEOS \cite{macfarlane06}, an EOS and opacity code with detailed configuration accounting.  Of particular interest is the value of $\gamma$ in
the xenon liner over the range of plasma-liner parameters spanning $r_m$ to $r_{\rm stag}$.
Using the non-LTE EOS table,
we evaluate $\gamma$ for xenon using the relationship \cite{zeldovich}
\begin{equation}
\label{eq:gamma}
\gamma=\frac{p}{\rho \epsilon} + 1,
\end{equation}
where $\epsilon$ is the total internal energy including both thermal and ionization/excitation energies,
and show the results in Fig.~\ref{fig:gamma-xe}. Equation~(\ref{eq:gamma}) shows that
a larger $\epsilon$ for given $p$ and $\rho$ will bring $\gamma$ closer to unity (i.e., below
$\gamma=5/3$ of an ideal gas); a decrease of $\gamma$ below 5/3 is
a measure of the extra energy sinks due to ionization and atomic excitation
(and associated line radiation) over that of an ideal gas.
Figure~\ref{fig:gamma-xe} shows that for the entire range of $T$ (assuming $T_e=T_i$)
and $n_i$, and in particular
for the most-relevant range of $T\sim 0.1$--10~eV during plasma-liner convergence,
$\gamma \lesssim 1.25$.  This is desirable in order
to keep the plasma liner
cold and to maintain a high $M$ during convergence, as discussed in Sec.~\ref{sec:mach}.
The same EOS/opacity tables are also used as input to 1D radiation-hydrodynamic
simulations of spherically imploding plasma liners, which are described in Sec.~\ref{sec:simulations}.
Although the opacity [Fig.~\ref{fig:opacity-cooling-Xe}(a)] and radiative-cooling
[Fig.~\ref{fig:opacity-cooling-Xe}(b)] data are not used explicitly in the analyses in this paper, they
are used in the simulation of a plasma-liner implosion in Sec.~\ref{sec:simulations} and are
included here for reference.

\subsection{Hydrodynamic properties}
\label{sec:hydro-calcs}

\subsubsection{Mach number}
\label{sec:mach}

\begin{figure}
\centerline{\includegraphics[width=2.8truein]{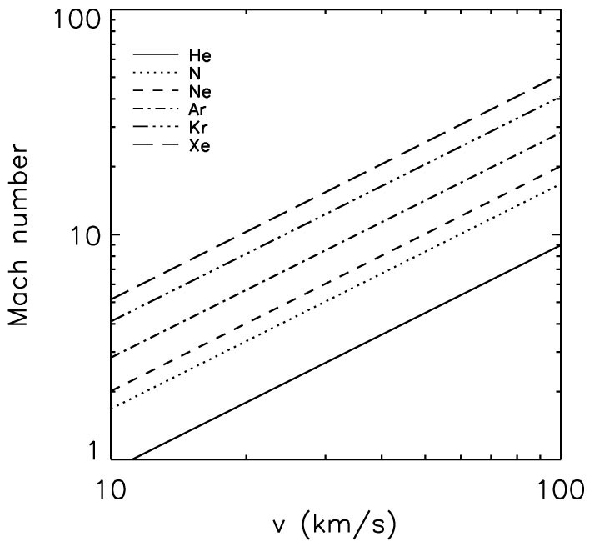}}
\caption{\label{fig:mach}Mach number $M$ versus liner implosion speed for different liner species,
assuming $T_e=T_i=2$~eV, $Z=1$, and $\gamma=1.3$.}%
\end{figure}

Because the peak liner ram pressure is expected to scale as $M^{3/2}$ \cite{awe11,davis12},
it is desirable to achieve high $M$, which will also reduce the amount of jet spreading and density degradation as the jets 
propagate from the guns to $r_m$ \cite{langendorf17pop}.  Achieving $M \ge 10$ in a subscale experiment is a reasonable criterion 
to be in a physics regime that is relevant to a reactor-relevant plasma liner.
Figure~\ref{fig:mach} shows $M$ (as defined in Sec.~\ref{sec:calcs}) versus liner implosion speed
for different species.  For helium, $v\gtrsim 100$~km/s
is needed for $M=10$, but for nitrogen (and heavier elements), $v\gtrsim 50$~km/s (or lower
for heavier elements) is sufficient.

\subsubsection{Simulation}
\label{sec:simulations}

\begin{figure}
\centerline{\includegraphics[width=2.8truein]{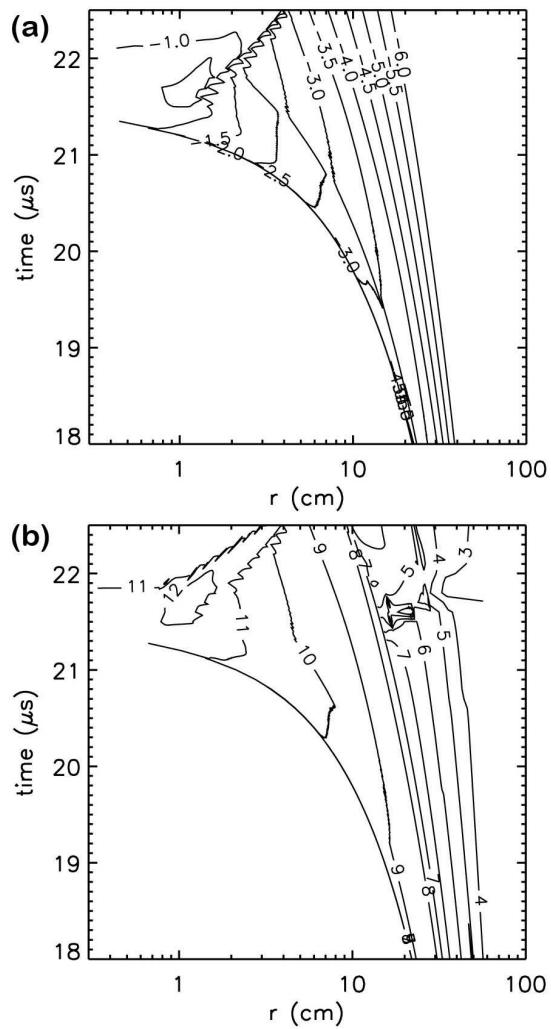}}
\caption{\label{fig:rho-pram-Xe}Contours of constant (a)~$\log(\rho~\mbox{[g/cc]})$ 
and (b)~$\log(\rho v^2 \mbox{[Pa]})$ vs.\ time and radius for
a PJMIF reactor-relevant, imploding xenon plasma liner, from a HELIOS simulation.
Peak $\rho=0.72$~g/cc and $\rho v^2=26.7$~Mbar are reached at $t=21.6$~$\mu$s and $r = 1.16$~cm.}
\end{figure}

To develop knowledge of the plasma-liner parameters from $r_m$ to $r_{\rm stag}$, we used HELIOS \cite{macfarlane06}, a 1D radiation-hydrodynamic code,
to simulate an imploding 1.62-cm-thick xenon plasma liner, starting at $r_m=150$~cm with
spatially uniform $v=70$~km/s, $n_0=10^{17}$~cm$^{-3}$, and $T_e=T_i=2$~eV\@.
Note that this simulation does not include a magnetized target; it is a xenon liner imploding on vacuum.
The simulation uses non-LTE xenon EOS/opacity data [Figs.~\ref{fig:zbar-Xe}(a), \ref{fig:pressure-energy-Xe}, and \ref{fig:opacity-cooling-Xe}] generated using PROPACEOS,
includes thermal and radiation transport, and allows $T_e$ and $T_i$ to evolve separately.
The liner is modeled using 300 computational zones (i.e., an average of 54~$\mu$m/zone at the initial 
time step).
Figure~\ref{fig:rho-pram-Xe} shows the temporal and spatial evolution of 
$\rho$ and $\rho v^2$, the peak values of
which are 0.72~g/cc and 26.7~Mbar, respectively,
at $t=21.6$~$\mu$s and $r=1.16$~cm.  Optimization of liner parameters and 
profiles, e.g., see \cite{kagan11}, is
needed in order to achieve the desired peak ram pressure of
150~Mbar at the same initial kinetic energy.

\section{Physics criteria to be satisfied in a subscale plasma liner experiment}
\label{sec:criteria}

The results in Sec.~\ref{sec:calcs} guide us to the important physics criteria that must be satisfied in
a subscale plasma liner experiment to be of relevance to the full-scale reactor-relevant plasma liner.
We state them as follows:
\begin{itemize}
\item $M>10$ at $r_m$
\item plasma must be collisional for the entire duration from $r_m$ to $r_{\rm stag}$, i.e., $\tau_i \ll 
\tau_{\rm
implosion}$
\item effective $\gamma$ of the liner species should be $<5/3$ as is the case for xenon.
\end{itemize}

\section{Requirements for a subscale plasma liner experiment}
\label{sec:requirements}

We use Eq.~(\ref{eq:mass}) to estimate $M$ and $E_0$ of a
subscale plasma liner experiment that satisfies the physics criteria stated in Sec.~\ref{sec:criteria}.
Rather than fixing $\rho_{\rm stag} r_{\rm stag}^2$ as we did for the PJMIF-reactor scenario, 
we use the criteria stated in Sec.~\ref{sec:criteria} and the constraints of the PLX vacuum chamber and
near-term plasma guns to set $r_m$, $L$, $n_0$, $v$, and $m$,
picking the lowest allowed values for each quantity in order to minimize $E_0$ 
and $M_{\rm liner}$ of a subscale experiment.

\begin{figure}
\centerline{\includegraphics[width=2.7truein]{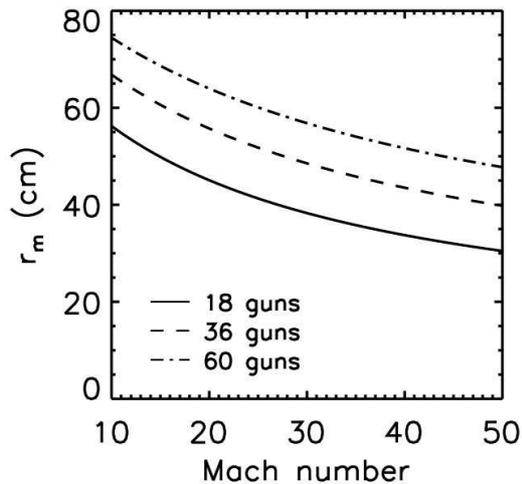}}
\caption{\label{fig:rm-vs-M}Jet-merging radius $r_m$ vs.\ $M$ for different numbers of plasma jets,
assuming $r_{j0}=5$~cm, $r_w=110$~cm, and $\gamma=1.3$.}
\end{figure}

First, we determine $r_m$, which is given by (and plotted in Fig.~\ref{fig:rm-vs-M}) \cite{cassibry13}
\begin{equation}
r_m = \frac{r_{j0}[M_j (\gamma - 1)/2 + 1] + r_w}{1+(2/N^{1/2})[M_j (\gamma -1)/2+1 ]},
\label{eq:rm}
\end{equation}
where $r_{j0}$ is the initial jet radius, $M_j$ the initial jet Mach number, $r_w$ the chamber-wall radius 
(where jets are launched), and $N$ the number of jets.  Equation~(\ref{eq:rm}) includes the effect
of jet spreading at the maximum rate $2C_s/(\gamma-1)$, where $C_s$ is the sound speed, 
and thus provides an upper bound on $r_m$.  We choose $r_m=75$~cm as an upper bound, corresponding to
$N=60$ and $M=10$ (see Fig.~\ref{fig:rm-vs-M}).  Lower $N$ and higher $M$ will result in lower $r_m$
and therefore lower $E_0$.

To determine $L$, we assume that the initial jet length equals $L$ at $r_m$.  
Upgrades to pulsed coaxial guns \cite{witherspoon09} for the subscale
plasma-liner-formation experiments were expected to achieve $L=10$~cm
(and indeed recently have done so \cite{case17aps}), and thus we use $L=10$~cm.

The minimum $n_0$ at $r_m$ is determined based on satisfying the 
collisionality requirement $\tau_i \ll \tau_{\rm implosion}$, i.e.,
\begin{eqnarray*}
\frac{\tau_i}{\tau_{\rm implosion}} & = &\frac{(4.80\times 10^{-8} Z^4 \mu^{-1/2} n_i \log\Lambda T_i^{-3/2})^{-1}}{r_m/v} \ll 1 \\
\rightarrow n_i & \gg & \frac{\mu^{1/2}T_i^{3/2}v}{4.80\times 10^{-8}Z^4\log\Lambda r_m}\approx 2
\times 10^{12}~\mbox{cm}^{-3},
\end{eqnarray*}
where we have used $\mu=14.01$ (nitrogen), $T=2$~eV,
$v=50$~km/s, $Z=1$, $\log\Lambda=7$, and $r_m=75$~cm.
Thus, we set $n_0>10^{14}$ to strongly satisfy the collisionality requirement.  Note that if $T$ rises
during liner convergence, the minimum $n_0$ requirement will decrease due to the $T_i^{3/2}/Z^4
\sim T_i^{-1/2}$ scaling (where, again, $Z\sim T^{1/2}$ is used as a simple model for ionization).

For $v$, we refer to Fig.~\ref{fig:mach} to see that we must have $v\approx 50$~km/s to have $M
\approx 10$ for nitrogen, which is the lowest-mass element being considered that will allow us to get 
$M=10$ while staying well below $v=100$~km/s.

\begin{figure}
\centerline{\includegraphics[width=2.8truein]{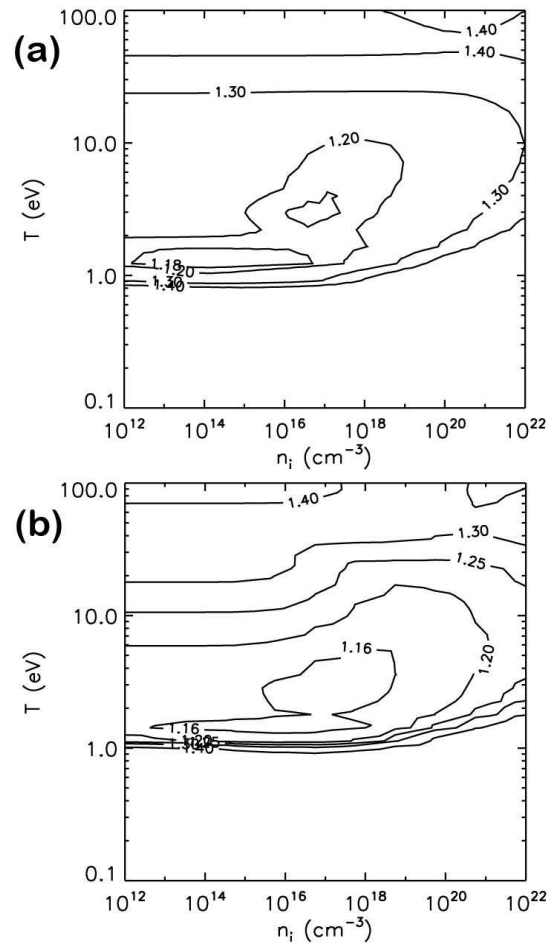}}
\caption{\label{fig:gamma-N} Contours of polytropic index $\gamma$
as a function of $T$ ($=T_e=T_i$) and $n_i$ for (a)~nitrogen
and (b)~argon, 
calculated using Eq.~(\ref{eq:gamma}) and non-LTE EOS tables from PROPACEOS\@.}
\end{figure}

Finally, using Eq.~(\ref{eq:gamma}) and a PROPACEOS non-LTE nitrogen EOS, 
we verify that $\gamma< 5/3$ for a nitrogen plasma liner so as to have
the desirable energy sink from ionization and atomic excitation to keep $M$ high,
as exhibited by xenon (Fig.~\ref{fig:gamma-xe}).  Figure~\ref{fig:gamma-N}(a) shows
that a nitrogen plasma liner in a subscale experiment would have
$\gamma<1.3$ for the most-relevant range of $T\sim 1$--10~eV\@.
Argon, which is also being used extensively in the subscale experiment, has $\gamma < 1.4$ for the same range [Fig.~\ref{fig:gamma-N}(b)].

With the values $r_m=75$~cm, $L=10$~cm, $n_0=10^{14}$~cm$^{-3}$, 
$v=50$~km/s, and nitrogen as the liner
species, Eq.~(\ref{eq:mass}) tells us that $E_0=2.1$~kJ and $M_{\rm liner}=1.65$~mg.  
To obtain scaling data in the subscale
experiment with species up to xenon would require nearly a factor of ten higher
mass (15.4~mg) and energy (19.7~kJ)\@.  To further scale up $n_0$ an order of magnitude
or $v$ up by a factor of three (or some combination thereof)
would require another factor of ten higher in energy, resulting in
$E_0\approx 200$~kJ for the ideal subscale plasma-liner-formation experiment, capable
of a decade in energy scaling for the heaviest element xenon.

Due to budgetary constraints, the subscale plasma-liner-formation experiment now being developed
\cite{hsu18} is expected to culminate with $N = 36$ guns and total capacitor stored energy
of $E_{\rm cap}=260$~kJ\@.  For $M=10$ and $N=36$, Fig.~\ref{fig:rm-vs-M} shows that $r_m \approx 65$~cm.  
Using $v=50$~km/s and $L=10$~cm, Eq.~(\ref{eq:mass}) gives $E_0=14.5$~kJ for xenon.
If the gun electrical efficiency is $\eta =0.25$ (as suggested by the ongoing
work \cite{case17aps,hsu18}), then the maximum liner energy 
is $\eta E_{\rm cap} = 65$~kJ,
meaning there is still a
reasonable factor of 4.5 headroom above $E_0=14.5$~kJ (xenon) for scaling studies.
Lower-mass species would have a higher headroom in energy for scaling studies.
\section{Summary}
\label{sec:summary}

We laid out the key requirements of a subscale plasma-liner-formation experiment to produce
data of direct relevance for addressing key scientific issues of a full-scale, reactor-relevant
plasma liner.  The key scientific issues of plasma-liner formation via merging hypersonic plasma jets
being addressed in the near term are:  (1)~determining the scaling of peak liner ram pressure with
initial plasma-jet parameters, and (2)~assessing
liner non-uniformity and prospects for non-uniformity control.

To ensure that a subscale plasma-liner-formation experiment can address the above issues in a way
that is relevant to the full-scale plasma liner, we identified key physics criteria that must be satisfied in the subscale experiment.  These criteria are: (1)~the plasma liner must be very collisional all the way
from the merging radius to stagnation, (2)~the liner Mach number must be high, i.e., $M>10$, (3)~the liner species must satisfy $M>10$ while possessing the
EOS and radiative properties, i.e., $\gamma < 5/3$, of a high-$Z$ species such as Xe, as
desired in a fusion-relevant liner.

A thirty-six-gun experiment with liner kinetic energy up to $E_0=65$~kJ that satisfies the physics criteria and requirements presented in this paper is now under development \cite{hsu18}.

\begin{acknowledgements}
The authors acknowledge many fruitful discussions with Drs.\ F. Douglas Witherspoon and Jason Cassibry, and also
the U.S. Department of Energy Advanced Research Projects Agency--Energy (ARPA-E), whose launch of the Accelerating Low-cost Plasma Heating and Assembly (ALPHA) Program motivated this work.
Strong Atomics, LLC, is acknowledged for supporting HyperJet Fusion Corporation's 
contributions to the subscale plasma liner experiment under development.
\end{acknowledgements}



\end{document}